# Molecular dynamics and optimization studies of horse prion protein wild type and its S167D mutant

**Jiapu Zhang**

Jiapu_zhang@hotmail.com, j.zhang@federation.edu.au

**Simple Summary:** Prion diseases or called transmissible spongiform encephalopathies (TSEs) are fatal neurodegenerative diseases. Prion disease has not been reported in horses up to now; therefore, horses are known to be a species resistant to prion diseases. Residue S167 in horse has been cited as critical protective residue for encoding PrP conformational stability in prion-resistance. This article is to do molecular dynamics and optimization studies on the horse PrP wild-type and its S167D mutant respectively to understand their conformational dynamics and optimized conformation.

**Abstract:** Prion diseases or called transmissible spongiform encephalopathies (TSEs) are fatal neurodegenerative diseases characterised by the accumulation of an abnormal prion protein isoform ($PrP^{Sc}$: rich in β-sheets - about 30% $\alpha$-helix and 43% β-sheet), which is converted from the normal prion protein ($PrP^{C}$: predominant in $\alpha$-helix - about 42% $\alpha$-helix and 3% β-sheet). However, prion disease has not been reported in horses up to now; therefore, horses are known to be a species resistant to prion diseases. Residue S167 in horse has been cited as critical protective residue for encoding PrP conformational stability in prion-resistance. According to the "protein-only" hypothesis, $PrP^{Sc}$ is responsible for both spongiform degeneration of the brain and disease transmissibility. Thus, understanding the conformational dynamics of $PrP^{Sc}$ from $PrP^{C}$ is a key to developing effective therapies. This article is to do molecular dynamics and optimization studies on the horse PrP wild-type and its S167D mutant respectively to understand their conformational dynamics and optimized confirmation; interesting results will be discussed.

## Introduction

Prion diseases are incurable neurodegenerative diseases caused by aberrant conformations of the prion protein (PrP). Many animals develop similar diseases, but rabbits, dogs, and horses show unusual resistance to prion diseases [1-15]. This resistance could be due to protective changes in the sequence of the corresponding PrP in each animal. Structural studies have identified S174, S167, and D159 as the key protective residues in rabbit, horse, and dog PrP, respectively [1-15]. But no systemic molecular dynamics (MD) studies currently support the protective activity of these residues, especially for horse PrP residue S167. Experimental laboratory results revealed that expression of horse PrP-S167D (which carries substitution for the equivalent residue in hamster (the species being susceptible to prion diseases) PrP) shows high toxicity in behavioural and anatomical assays [14]. Thus, this article aims to MD study horse PrP wild-type (WT) NMR structure 2KU4.pdb and to optimization study S167D-mutant (Mutant) homology structure (constructed by this article). We will present in this article useful protective bioinformatics of S167 and discuss the structural features that make horse PrP more stable. The findings of this article might contribute to the development of drugs/compounds that stabilize PrP structure and prevent the formation of toxic conformations of prion diseases.

Here we detail the central topic on PrP more in this introduction section. Unlike bacteria and viruses which are based on DNA and RNA, prions are unique as disease-causing agents since they are misfolded proteins. Prion contains no nucleic acids and it is a misshapen or confirmation-changed protein that acts like an infectious agent. Prion diseases are called "protein structural conformational" diseases. Normal prion protein is denoted as $PrP^{C}$ and diseased infectious prion is denoted as $PrP^{Sc}$. $PrP^{C}$ is predominant in $\alpha$-helices but $PrP^{Sc}$ are rich in β-sheets in the form as amyloid fibrils. $PrP^{C}$ is normal protein found on the membranes of cell, including several blood components of which platelets constitute the largest reservoir in humans. Several topological forms exist; one cell surface form anchored via glycolipid and two transmembrane forms. The normal protein has a complex function, which continues to be investigated by now; the cleavage of $PrP^C$ in peripheral nerves causes the activation of myelin repair in Schwann cells, $PrP^C$ regulates cell death, $PrP^C$ may have a function in maintenance of long-term memory, $PrP^C$ may play roles in innate immunity and stem cell renewal, etc. $PrP^C$ binds $Cu^{2+}$ ions with high affinity; the significance of this property is not clear but it is presumed to relate to the protein's structure or function. $PrP^C$ is not sedimentable, meaning it cannot be separated by centrifuging techniques. $PrP^C$ is readily digested by proteinase K and can be liberated from the cell surface by the enzyme phosphainositide phospholipase C, which cleaves the glycophosphatidylinositol glycoplipid anchor. $PrP^C$ plays an important role in cell-cell adhesion and intracellular signaling in vivo, and may therefore be involved in cell-cell communication in the brain. $PrP^{Sc}$ always causes prion disease. Several highly infectious, brain-derived $PrP^{Sc}$ structures have been discovered by cryo-EM; another brain-derived fibril structure isolated from humans with prion disease GSS syndrome has also been determined. Often $PrP^{Sc}$ is bound to cellular membranes, presumably via its array of glycolipid anchors; however, sometimes the fibers are dissociated from membranes and accumulate outside of cells in the form of plaques. S167 in $PrP^C$ is a protective residue and generates a more compact and stable structure in the C-terminal subdomain of the $PrP^C$ global domain [17, 15].

**Materials and Methods**

The material is same as the one of [13]: the NMR structure 2KU4.pdb of horse PrP WT (119-231). Basing on 2KU4.pdb, we just make one mutation S167D only at position 167 from hamster PrP residue ASP167 and get a homology structure for S167D-mutant of horse PrP. The optimization-study methods for the S167D-mutant are referred from [18]; i.e. taking into account the three-body movement we use the hybrid local search optimization method.

For both the WT and the Mutant, the low pH in the MD simulations is achieved by the change of residues HIS, ASP, GLU into HIP, ASH, GLH respectively and the Cl- ions added by the XLEaP module of AMBER package, and the neutral pH in the MD simulations is achieved by the change of residues HIS into HID and the $Na^+$ ions added by the XLEaP module of AMBER package.

The MD simulations are done at 300 K, 350 K and 450 K respectively as Section 1.2 of [16] described; in order to make convenient for readers and for reproducibility, we list it as (I)-(III) as follows. *(I) At 300 K*: MD simulations used the ff03 force field of the AMBER packages (ambermd.org), in neutral or low pH environments (where residues HIS were changed into HID by the XLEaP module of AMBER packages in order to get the neutral pH environment). The systems were surrounded with a 12 ˚A layer of TIP3PBOX water molecules and neutralized by sodium (Na+ or Cl−) ions using XLEaP module of AMBER packages. The solvated proteins with their counter-ions were minimized mainly by the steepest descent method and then a small number of conjugate gradient steps were performed on the data, in order to remove bad hydrogen bond (HB) etc. contacts. Then the solvated proteins were heated from 100 to 300 K during 300 ps and then kept at 300 K for 700 ps, both in the constant NVT ensemble using Langevin thermostat algorithm. The SHAKE algorithm and PMEMD algorithm with nonbonded cut-offs of 12 ˚A were used during the heating. Enough equilibrations had been done in the constant NPT ensemble under Berendsen thermostat during 5 ns until the RMSD, PRESS, VOLUME and DENSITY were sufficiently stable, where the RMSD values did not fluctuate very much within a few picoseconds. After equilibrations, production MD phase was carried out at 300 K for 30 ns using constant pressure and temperature NPT ensemble and the PMEMD algorithm with nonbonded cutoffs of 12 ˚A during the production simulations. Step size for equilibration is 0.5 fs, and 1 fs is for the production runs. The structures were saved to file every 1000 steps. The MD was finished in all in 36 ns. *(II) At 350 K:* 350 K might be a practical temperature for many experimental laboratory works. MD simulations used the ff03 force field of the AMBER packages, in neutral and low pH environments (where residues HIS, ASP, GLU were changed into HIP, ASH, GLH respectively by the XLEaP module of AMBER packages in order to get the low pH environment). The systems were surrounded with a 12 ˚A layer of TIP3PBOX water molecules and neutralized by sodium ions using XLEaP module of AMBER packages. The solvated proteins with their couterions were minimized mainly by the steepest descent method and then a small number of conjugate gradient steps were performed on the data, in order to remove bad contacts. Then the solvated proteins were heated from 100 to 300 K during 1 ns (with step size 1 fs) and from 300 to 350 K during 1 ns (with step size 2 fs). The thermostat algorithm used is the Langevin thermostat algorithm in constant NVT ensembles. The SHAKE algorithm and PMEMD algorithm with nonbonded cutoffs of 12 ˚A were used during the 2 ns' heating. Equilibrations were done in constant NPT ensembles under Langevin thermostat for 2 ns. After equilibrations, production MD phase was carried out at 350 K for 30 ns using constant pressure and temperature NPT ensemble and the PMEMD algorithm with nonbonded cutoffs of 12 ˚A during simulations. Step size for the production runs is 2 fs. The structures were saved to file every 1000 steps. The MD was finished in all in 34 ns. *(III) At 450K:* MD simulations used the ff03 force field of the AMBER packages, in neutral and low pH environments. The systems were surrounded with a 12 ˚A layer of TIP3PBOX water molecules and neutralized by sodium ions using XLEaP module of AMBER packages. The solvated proteins with their counterions were minimized mainly by the steepest descent method and then a small number of conjugate gradient steps were performed on the data, in order to remove bad contacts. Then the solvated proteins were heated from 100 to 450 K step by step during 3 ns. The thermostat algorithm used is the Langevin thermostat algorithm in constant NVT ensembles. The SHAKE algorithm and PMEMD algorithm with nonbonded cutoffs of 12 ˚A were used during the heating. Equilibrations were done in constant NPT ensembles under Langevin thermostat for 5 ns. After equilibrations, production MD phase was carried out at 450 K for 30 ns using constant pressure and temperature NPT ensemble and the PMEMD algorithm with nobonded cutoffs of 12 ˚A during production simulations. Step size for equilibration was 0.5 and 1 fs for the production runs. The structures were saved to file every 1000 steps. The MD was finished in all in 38 ns.
The study utilized computational approach to study the conformational stability of the target protein. For the mutant, the decision to use homology modeling rather than in silico mutagenesis, because currently the experimental structure of S167D-mutant is not produced by a NMR / X-ray / cryo-EM lab yet; the rationale behind selecting this method is high - the author's experiences show us that for a molecular structure just making one mutation is usually much agreeing with the experimental laboratory NMR / X-ray / cryo-EM structure. However, for a molecular structure if at the same time two mutations are made, the homology structure usually is far from the laboratory structure so that it is not reliable to use the homology structure. This is a point we should highlight the real applications of this study.
We should also note that in this article three dynamic replicas were performed and the results are not the average of the dynamics.

**Results and Discussion**

For convenience, we give some acronyms. For WT horse normal cellular prion protein [12] (horse $PrP^C$ with PDB entry 2KU4), its structural region GLY119-SER231 consists of β-strand 1 (β1: MET129-ALA133), $\alpha$-helix 1 ($\alpha$1: ASP144-ARG151), $3_{10}$-helix 1 ($3_{10}$H1: MET154-ARG156), β-strand 2 (β2: GLN160-TYR163), $3_{10}$-helix 2 ($3_{10}$H2: VAL166-GLU168), $\alpha$-helix 2 ($\alpha$2: GLN172-THR192), $\alpha$-helix 3 ($\alpha$3: GLU200-ARG228), and the loops linking them each other. As we all know, the stability of a protein is maintained by its salt bridges, hydrogen bonds, hydrophobic contacts, van der Waals contacts and disulfide bonds (for PrP monomer there exists a disulfide bond (S-S) between CYS179 and CYS214) etc. to drive to be able to perform the biological function of the protein; we denote SBs, HBs, HYDs, vdWs for salt bridges, hydrogen bonds, hydrophobic interactions, and van der Waals contacts respectively. We denote amino acids (or residues) as 'aa'. Residue 167 is in the β2-$\alpha$2 loop of PrPs, and both Syrian hamsters and horses have a well-defined and highly ordered β2-$\alpha$2 loop.

| SBs | salt bridges |
|---|---|
| HBs | hydrogen bonds |
| HYDs | hydrophobic interactions |
| vdWs | van der Waals |
| aa | amino acid (or residue) |
| S-S | disulfide bond |

Firstly, we optimized the WT and Mutant structures and their backbone atoms RMSD (root mean square deviation) value is 0.495090 angstroms (Figure 1). We find the mutation made that (i) *for the secondary structures* in the segment TYR169-ASN171 (before the N-terminal of $\alpha$2) Turn is changed into Coil, in the segment LYS194-GLU196 (next to the C-terminal of $\alpha$2) Coil is changed into Turn, and extend the C-terminal $\alpha$1 from Bend-Turn structure in the segment GLU152-ASN153; (ii) *for the SBs* ASP147-ARG148 (in $\alpha$1), ASP147-ARG151 (in $\alpha$1), ASP178-HIS177 (in $\alpha$2), ASP202-ARG156 (linking $\alpha$3 - $3_{10}$H1), GLU152-ARG151 (linking $\alpha$1-$3_{10}$H1-loop - $\alpha$1), GLU168-ARG164 (linking $3_{10}$H2 - β2-$3_{10}$H2-loop), GLU200-LYS204 (in $\alpha$3), GLU211-ARG208 (in $\alpha$3), GLU221-LYS220 (in $\alpha$3) (except for ASP144-ARG148 (in $\alpha$1), GLU196-ARG156 (linking $\alpha$2-$\alpha$3-loop - $3_{10}$H1)) disappeared in the Mutant, (iii) *for the HBs* 12 HBs are less than in WT (see details in Table 1 – we illuminate the importance of HB ASH/ASP202.OD1-TYR149.OH of WT in Figure 2 though it is weak during the 30 ns' MDs but not existing in the optimized structure of the mutant), and (iv) *for the π-interactions* π-cations PHE141-ARG208.NH2$^+$ (linking β1-$\alpha$1-loop - $\alpha$3), HIS177-LYS173.NZ$^+$ (in $\alpha$2) disappeared in the Mutant. Residue 167 is in the β2-$\alpha$2-loop and the mutation S167D results in increased negative charges on the surface around the β2-$\alpha$2-loop region (ASP is a negatively charged residue) (Figure 3).

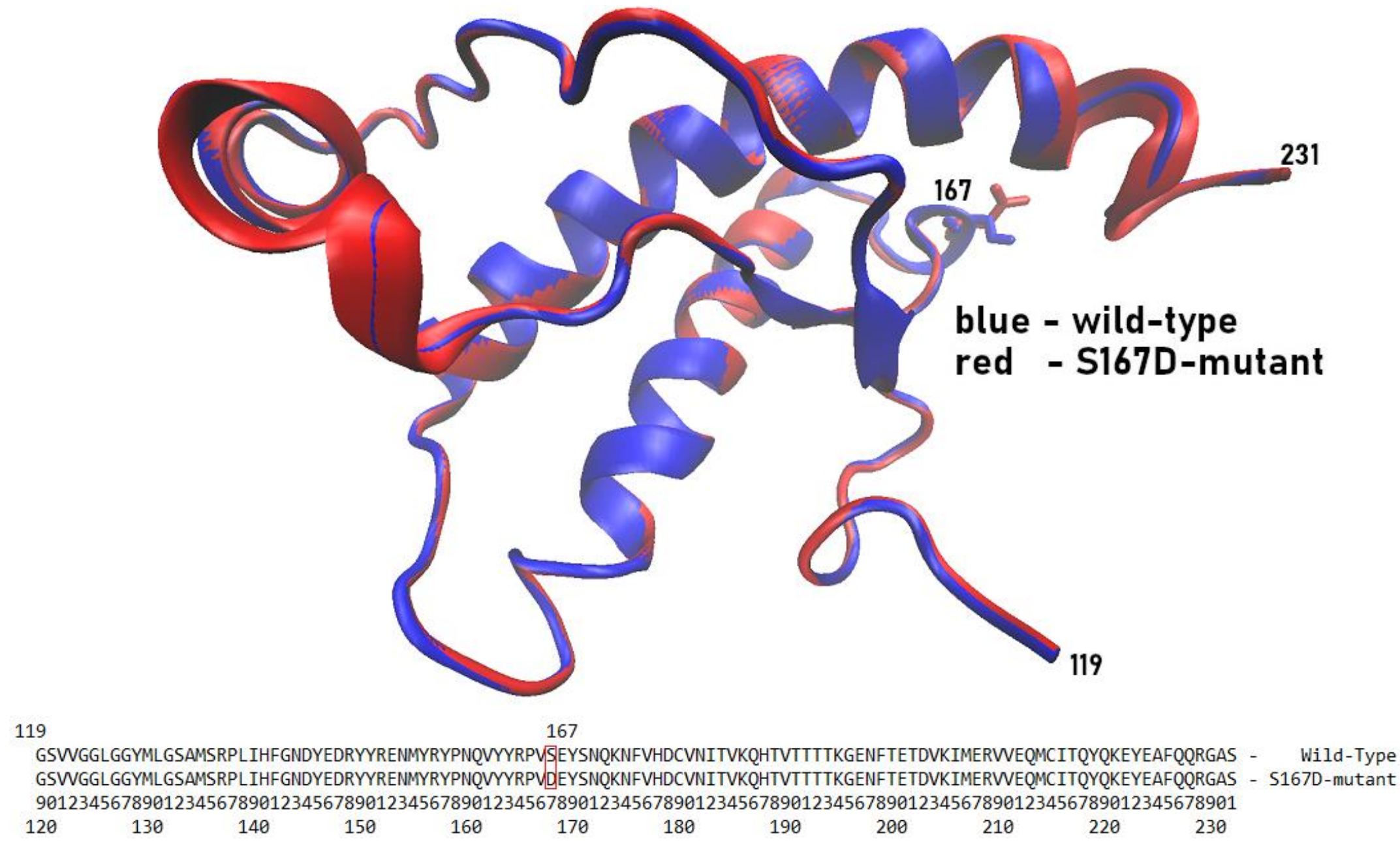

**Figure 1.** ***Performing a three-dimensional structural alignment (with RMSD value 0.495090 angstroms) and a primary sequence alignment of horse-PrP Wild-Type and its S167D-mutant.***

Distance between ASH/P202.OD1 and TYR149.OH during 30 ns

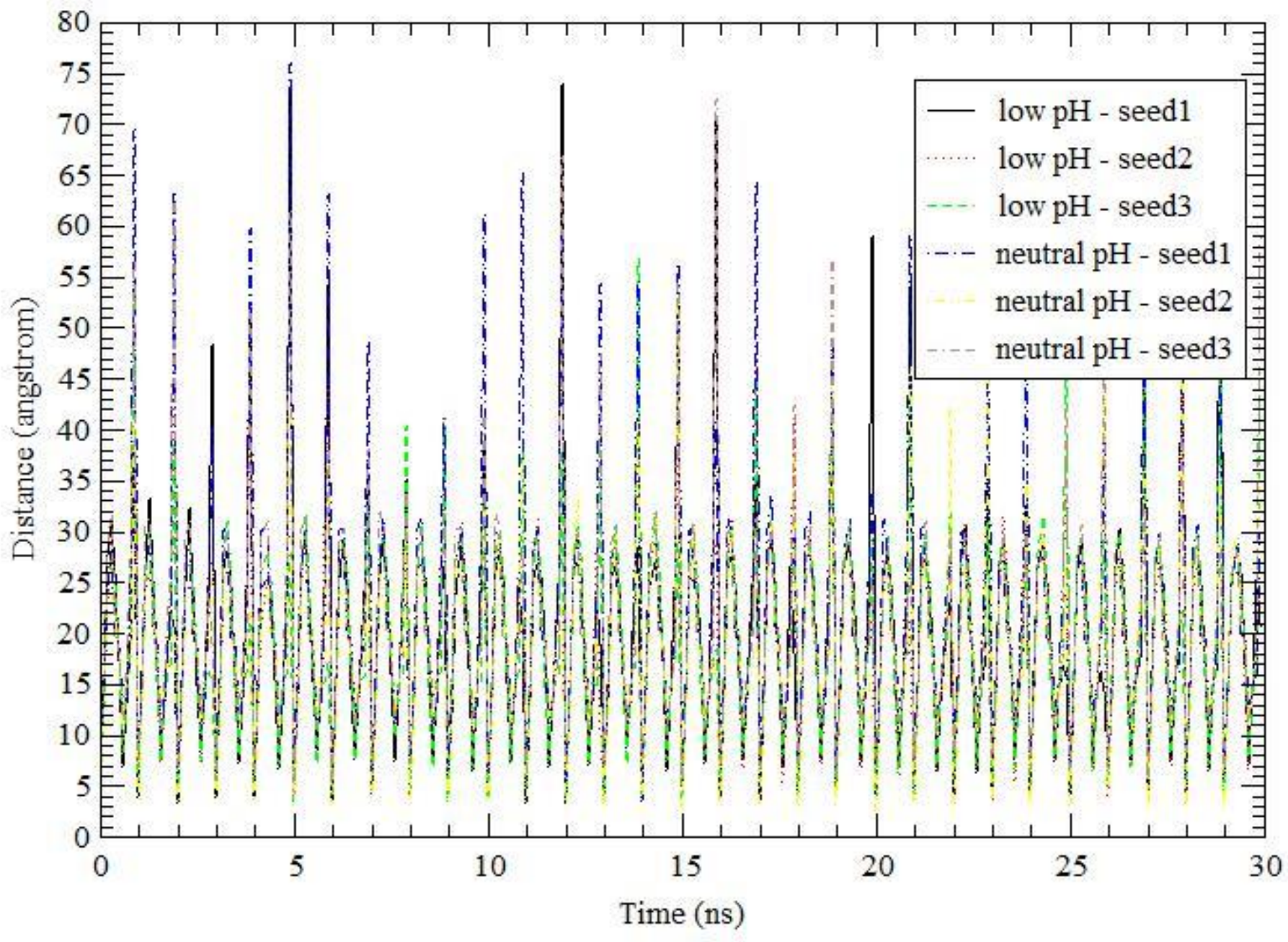

**Figure 2.** *The importance of HB ASH/ASP202.OD1-TYR149.OH of WT though it is weak during the 30 ns' MDs but not existing in the optimized structure of the S167D-mutant.*

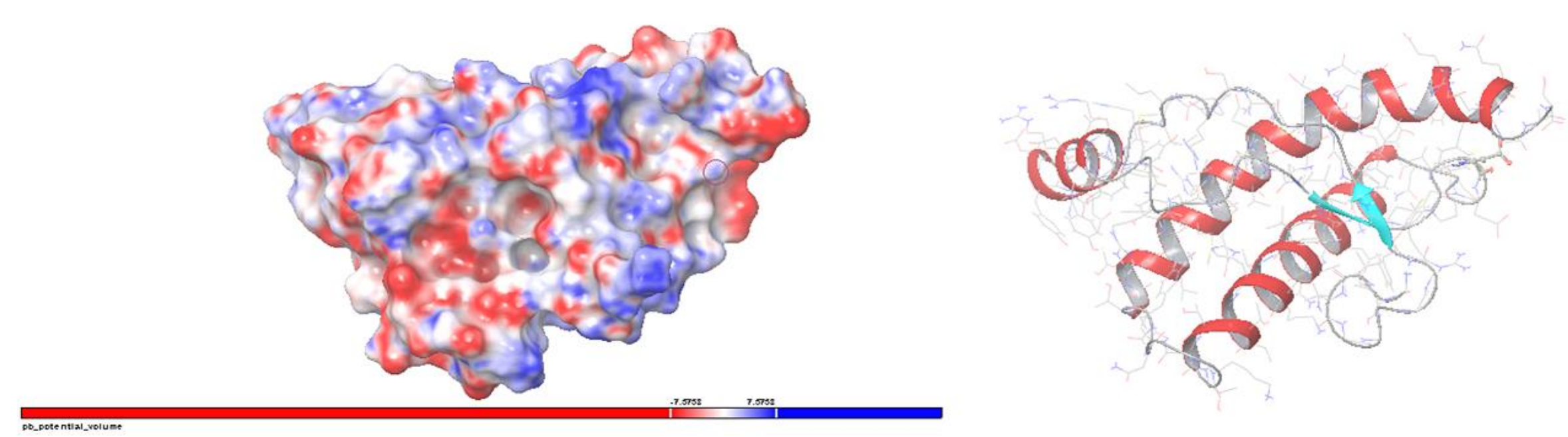

**Figure 3.** *Residue 167 is in the β2-α2-loop and the mutation S167D results in increased negative charges on the surface around the β2-α2-loop region (ASP is a negatively charged residue – blue-colored and circled in the Poisson-Boltzmann electrostatic potential surface); the right graph is the three-dimensional structure of S167D-mutant.*

**Table 1.** *The S167D mutation made changes in the network of hydrogen bonds (HBs):*

| ***HBs kept*** | *locations* | ***HBs lost*** | *locations* |
|---|---|---|---|
| ARG228.N-ALA224.O | in α3 | TYR149.OH-ASP202.OD1 | α1 - α3 |
| ASN181.N-HIS177.O | in α2 | ASN153.ND2-TYR149.O | α1-$3_{10}$H1-loop - α1 |
| CYS179.N-PHE175.O | in α2 | ARG156.N-ASN153.O | $3_{10}$H1 - α1-$3_{10}$H1-loop |
| CYS214.N-VAL210.O | in α3 | ARG164.NE-GLU168.OE2 | β2-$3_{10}$H2-loop - $3_{10}$H2 |
| GLN217.N-MET213.O | in α3 | ARG164.NH2-GLU168.OE1 | β2-$3_{10}$H2-loop - $3_{10}$H2 |
| GLU207.N-VAL203.O | in α3 | HIS177.ND1-ASP178.OD1 | in α2 |
| GLU211.N-GLU207.O | in α3 | ASP178.N-ASN174.O | in α2 |
| GLU221.N-GLN217.O | in α3 | HIS187.N-THR183.O | in α2 |
| GLY131.N-VAL161.O | β1 - β2 | THR192.N-THR188.O | in α2 |
| HIS177.N-LYS173.O | in α2 | LYS194.N-THR191.O | α2-α3-loop - α2 |
| ILE182.N-ASP178.O | in α2 | GLU196.N-THR191.OG1 | α2-α3-loop - α2 |
| ILE205.N-THR201.O | in α3 | PHE198.N-THR192.OG1 | α2-α3-loop - α2 |
| ILE215.N-GLU211.O | in α3 | GLN212.NE2-GLU211.OE2 | in α3 |
| LYS185.N-ASN181.O | in α2 | *GLN217.NE2-TYR163.OH* | α3 - β2 |
| LYS204.N-GLU200.O | in α3 | GLN219.NE2-ILE215.O | in α3 |
| MET129.N-TYR163.O | β1 - β2 | GLN227.NE2-GLU223.O | in α3 |
| MET134.N-ASN159.O | β1-α1-loop - $3_{10}$H1-β2-loop | SER231.OG-GLN227.O | C-terminal - α3 |
| MET206.N-ASP202.O | in α3 | | |
| MET213.N-VAL209.O | in α3 | ***new HBs*** | *locations* |
| PHE225.N-GLU221.O | in α3 | TYR150.N-GLU146.O | in α1 |
| THR183.OG1-CYS179.O | in α2 | ARG151.N-ASP147.O | in α1 |
| THR188.N-VAL184.O | in α2 | GLU152.N-ARG148.O | α1-$3_{10}$H1-loop - α1 |
| THR190.N-GLN186.O | in α2 | ASN171.ND2-ASN174.OD1 | $3_{10}$H2-α2-loop - α2 |
| THR216.N-GLN212.O | in α3 | GLN186.N-ILE182.O | in α2 |
| TYR149.N-TYR145.O | in α1 | | |
| TYR150.OH-PRO137.O | α1 - β1-α1-loop | | |
| TYR162.N-THR183.OG1 | β2 - α2 | | |
| TYR218.N-CYS214.O | in α3 | | |
| TYR222.N-TYR218.O | in α3 | | |
| VAL176.N-GLN172.O | in α2 | | |
| VAL180.N-VAL176.O | in α2 | | |
| VAL184.N-VAL180.O | in α2 | | |
| VAL189.N-LYS185.O | in α2 | | |
| VAL203.N-THR199.O | α3 - α2-α3-loop | | |
| VAL209.N-ILE205.O | in α3 | | |
| VAL210.N-MET206.O | in α3 | | |

Because the S167D mutation made some HBs & SBs in $\alpha 2$ & $\alpha 3$ and some $\pi$-cations lost, we may *see the molecular structure of the Mutant (compared with the WT) is unstable especially in the regions of $\alpha 2$ & $\alpha 3$ (especially at both terminals of $\alpha 2$).* On the contrary, in Section 4.1 of [16], the MD results show that *the N-terminal half of $\alpha 1$ is not stable at 350 K and 450 K, but the β2-α2-loop has less variation than other PrP loops* being due to the well-defined and highly ordered β2-α2-loop structure of horse PrP. These imply to us that the S167D mutation reversed the MD results of [16] – S167 is critical to contribution of horse PrP$^{C}$ structure stability.

The above is just a snapshot's results of WT or Mutant. Next, let us study 30ns' MD results of WT.

We first consider the secondary structure developments (Table 2 and Figures 4-6), HBs (Table 3), SBs (Table 4), and HYDs (Tables 5-6) results of 30ns' MDs of WT.

**Table 2.** ***Secondary Structures percentages of the WT at 300K, 350 K, 450 K during 30 ns' MD simulations (seed1, seed2, seed3) in neutral and low pH environments:***

| | | | **B β-bridge** | **G $3_{10}$-helix** | **I π-helix** | **H α-helix** | **E β-sheet** | **T Turn** | **S Bend** |
|---|---|---|---|---|---|---|---|---|---|
| *300 K* | *Low pH* | *Seed1* | 0.00% | 4.88% | 0.00% | 50.41% | 3.54% | 7.79% | 7.65% |
| | | *Seed2* | 0.26% | 3.32% | 0.09% | 48.54% | 4.63% | 8.34% | 9.00% |
| | | *Seed3* | 0.21% | 4.44% | 0.00% | 46.63% | 4.07% | 8.93% | 10.89% |
| | *Neutral pH* | *Seed1* | 0.00% | 4.53% | 0.00% | 51.20% | 3.54% | 5.04% | 8.73% |
| | | *Seed2* | 0.13% | 3.94% | 0.03% | 50.15% | 3.55% | 9.00% | 7.94% |
| | | *Seed3* | 0.50% | 2.90% | 0.00% | 51.17% | 3.73% | 8.34% | 8.83% |
| *350 K* | *Low pH* | *Seed1* | 0.21% | 2.85% | 0.00% | 49.39% | 3.42% | 8.34% | 10.18% |
| | | *Seed2* | 0.34% | 3.55% | 0.03% | 48.73% | 4.11% | 9.05% | 9.11% |
| | | *Seed3* | 0.37% | 3.24% | 0.00% | 49.29% | 4.14% | 8.67% | 8.20% |
| | *Neutral pH* | *Seed1* | 0.47% | 4.90% | 0.00% | 46.79% | 4.96% | 9.62% | 7.92% |
| | | *Seed2* | 0.18% | 3.60% | 0.03% | 49.18% | 3.49% | 9.04% | 9.60% |
| | | *Seed3* | 0.67% | 4.70% | 0.03% | 49.53% | 4.31% | 6.76% | 9.19% |
| *450 K* | *Low pH* | *Seed1* | 1.09% | 3.47% | 0.09% | 42.18% | 1.17% | 12.46% | 12.70% |
| | | *Seed2* | 0.93% | 4.85% | 0.12% | 41.37% | 3.60% | 14.31% | 12.84% |
| | | *Seed3* | 0.76% | 4.25% | 0.06% | 38.34% | 3.14% | 14.80% | 15.43% |
| | *Neutral pH* | *Seed1* | 0.81% | 4.35% | 0.00% | 38.38% | 0.64% | 16.43% | 14.61% |
| | | *Seed2* | 0.73% | 4.09% | 0.12% | 42.47% | 3.33% | 11.43% | 14.64% |
| | | *Seed3* | 0.28% | 4.07% | 0.09% | 45.13% | 4.05% | 10.64% | 7.31% |

Table 2 shows us compared with under neutral pH environment (i) at 300 K H $\alpha$-helices become less and E β-strands become more under low pH environment; (ii) at 350 K G $3_{10}$-hellices become less, for seed3 I $\pi$-helices become less, for seed2 & seed3 H $\alpha$-helices become less, and for seed2 E β-strands become more under low pH environment; and (iii) at 450 K B β-bridges become more, for seed1 G $3_{10}$-helices become less, for seed3 I $\pi$-helices become less, for seed2 & seed3 H $\alpha$-helices become less, and for seed1 & seed2 E β-strands become more under low pH environment. From Table 2, we can see the $PrP^C$ is predominant in $\alpha$-helix - at least 38.34% $\alpha$-helix, under neutral pH environment the percentages of $\alpha$-helix become less and less as the temperatures go up from 300 K to 350 K and to 450 K, however under low pH environment this rule is not obeyed for seed2 & seed3 from 300 K to 350 K. $PrP^C$ has β-sheet structure about 3%, and at 300 K the change from neutral pH environment to low pH environment makes the percentages of β-sheet increase (Table 2).

In Figure 4 we notice that at 300 K (i) under neutral pH environment, for seed2 during 0-11.8 ns the $\alpha$2 C-terminal segment aa 189-195 and for seed1 & seed3 the $\alpha$3 N-terminal segment aa 200-201 are in Turn structure; and (ii) under low pH environment, for seed2 during 0-9.4 ns segment aa 160-161 is extension of β1 sheet structure and for seed3 the $\alpha$3 C-terminal segment 220-226 is in Turn structure.

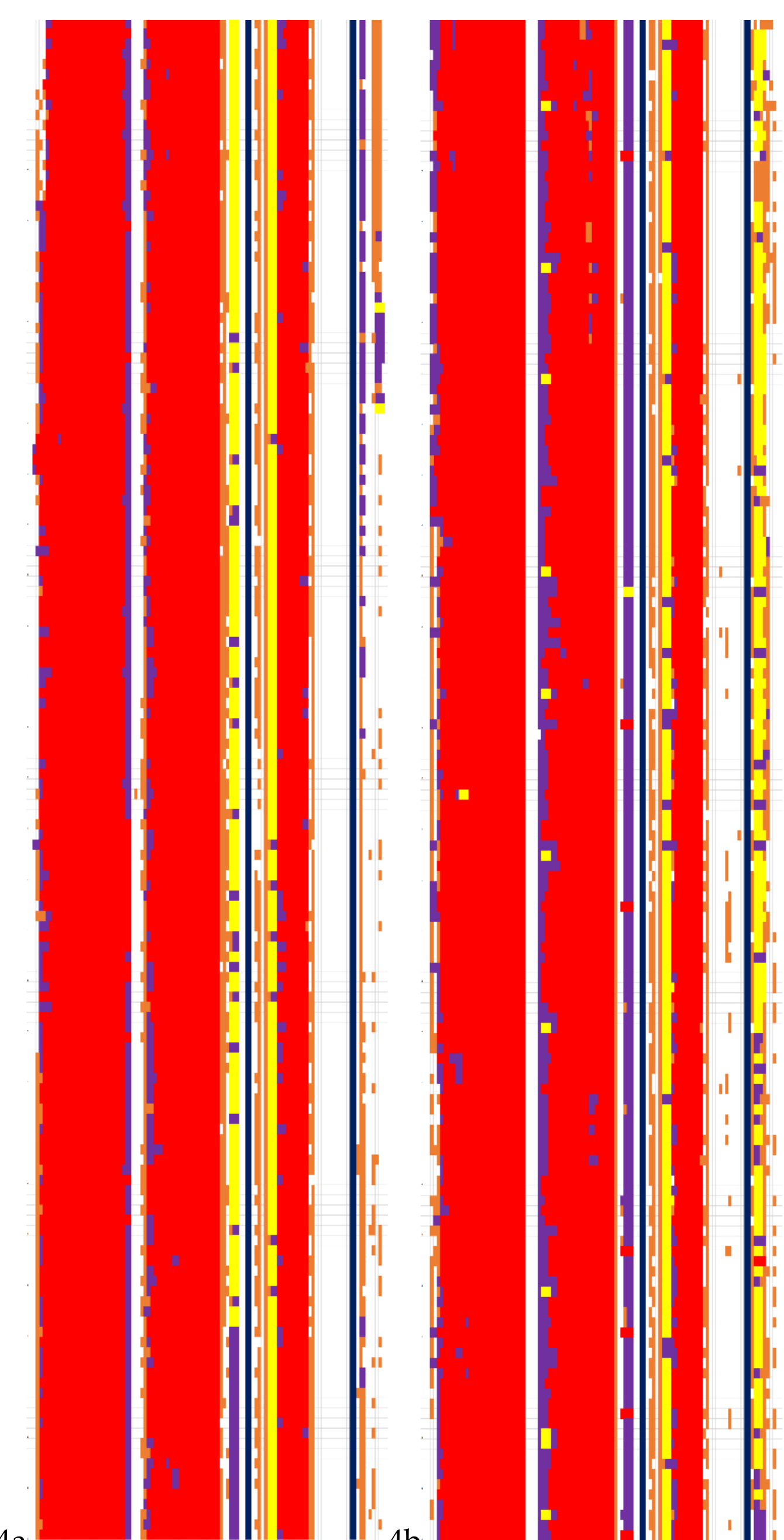

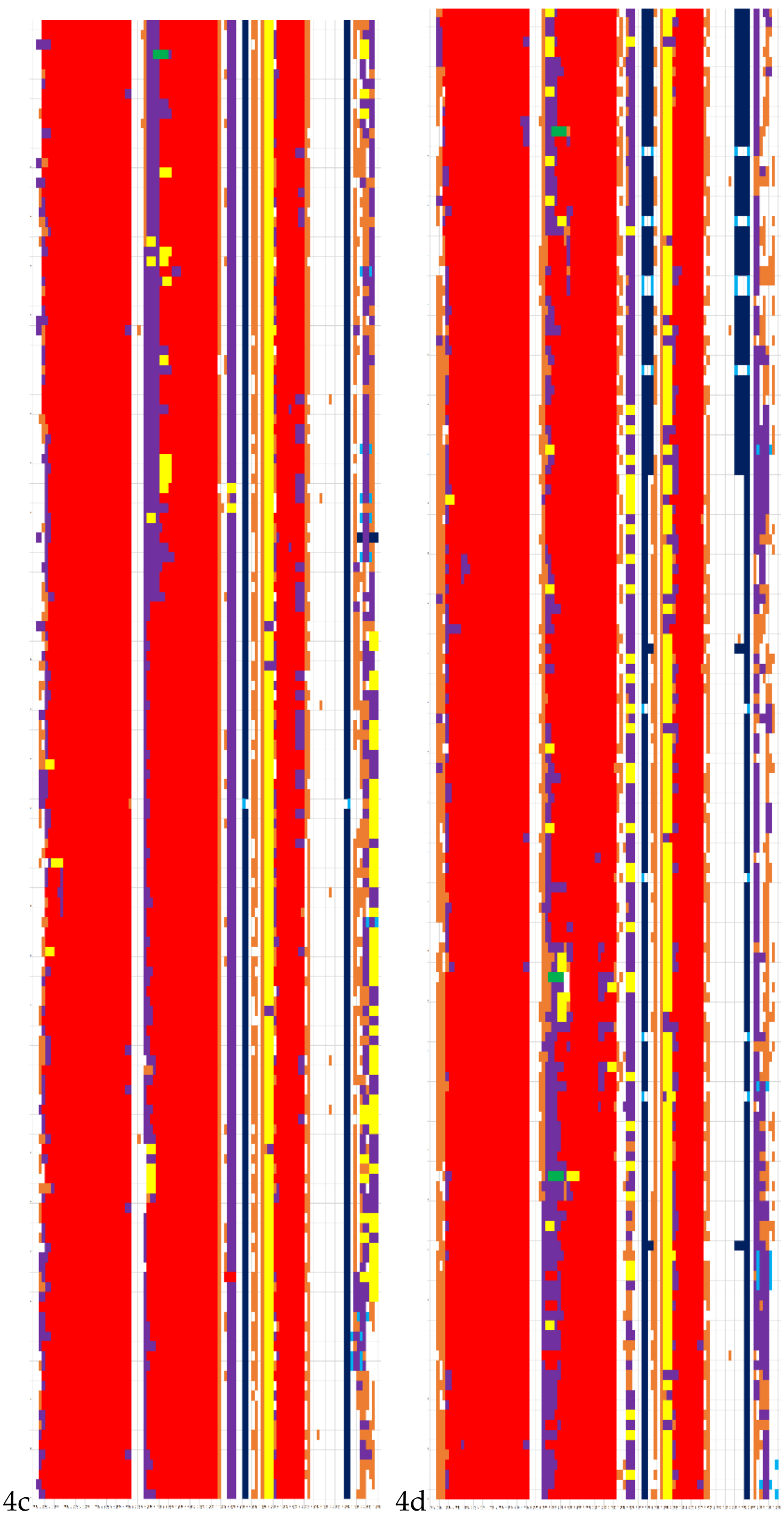

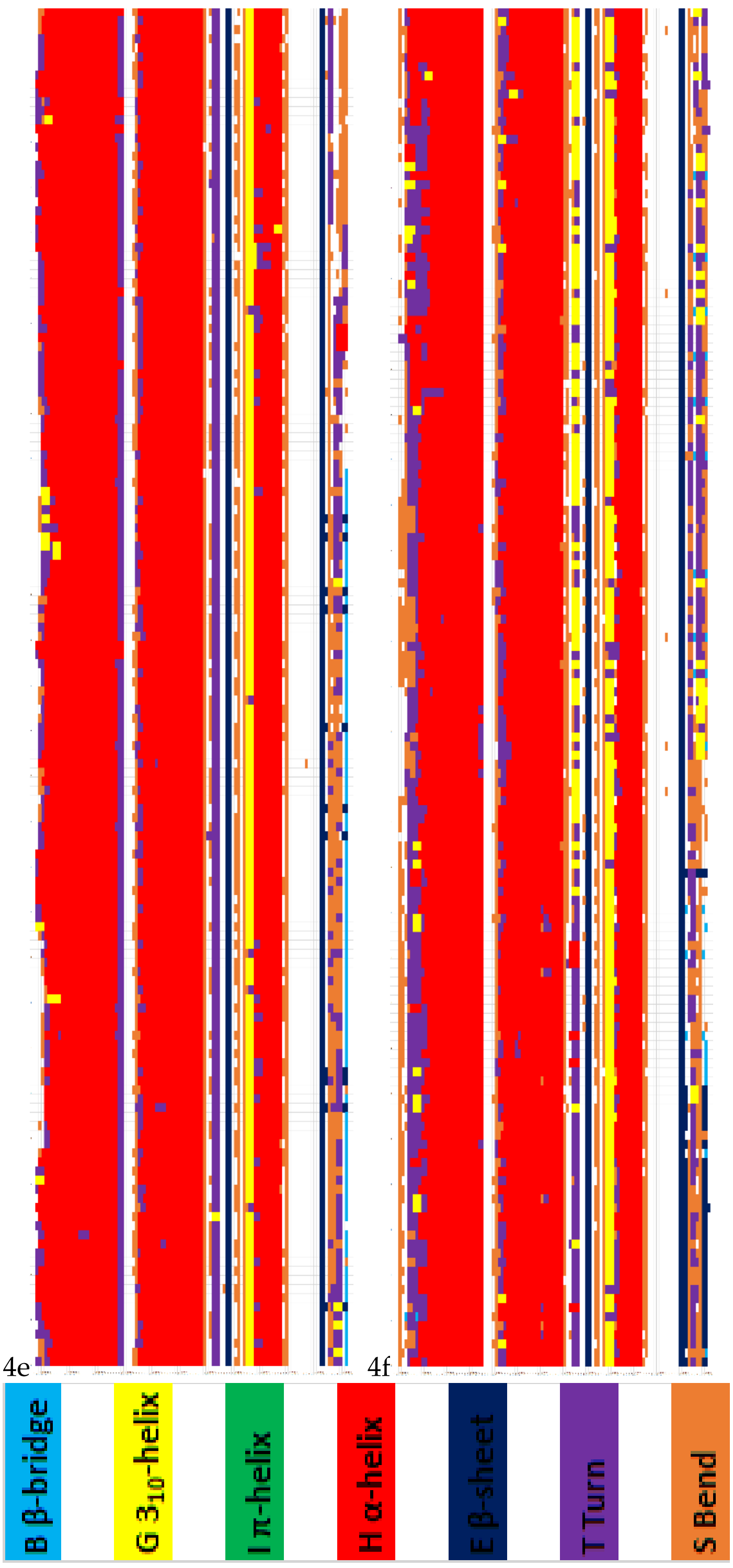

**Figure 4.** ***Secondary Structures of the WT at 300 K during 30 ns' MD simulations (up to down: seed1, seed2, seed3; left to right: neutral pH, low pH; 4a - neutral pH seed1, 4b - low pH seed1, 4c - neutral pH seed2, 4d - low pH seed2, 4e - neutral pH seed3, 4f - low pH seed3; for each of the six graphs its labels from up to down is 1 to 30 ns and its labels from right to left is PrP residue numbers 119-231). The meaning of each colours is also illuminated in the last graph.***

In Figure 5 we notice that at 350 K under neutral pH environment for seed1 (i) β1 E β-strand structure extends to segment aa 120-130 and (ii) $\alpha$1 H $\alpha$-helix structure unfolds in segment aa 152-157 during 13-30 ns.

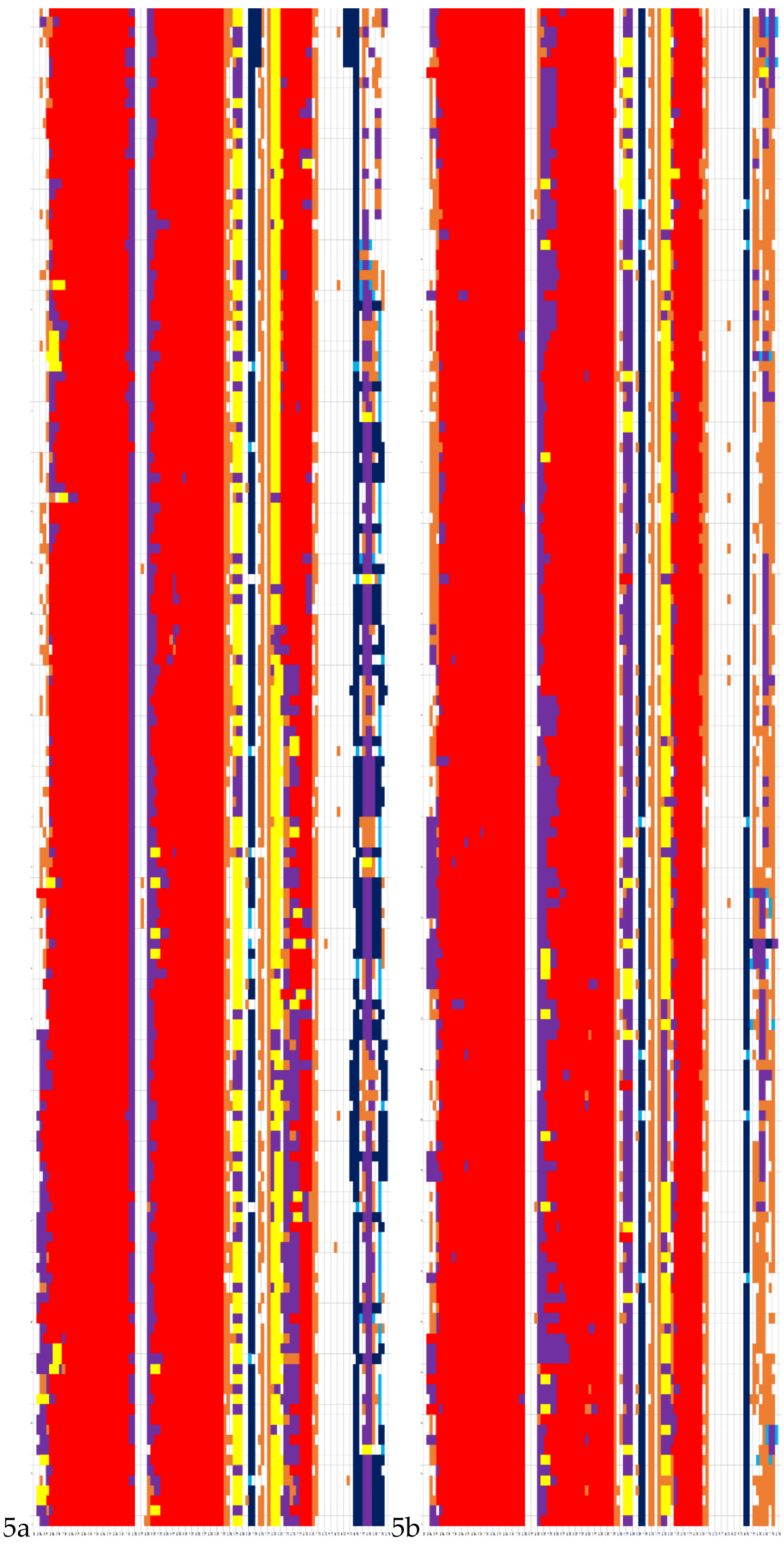

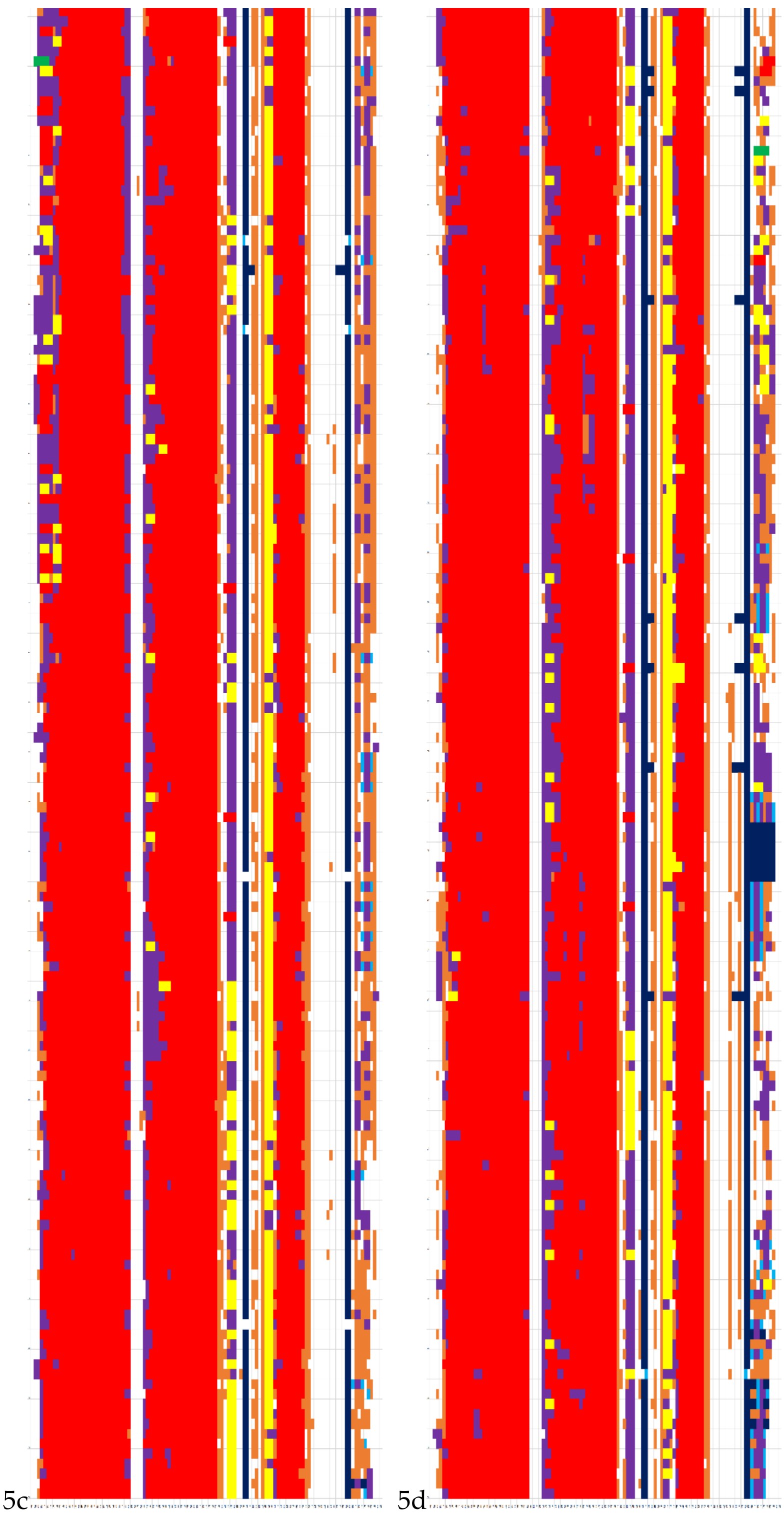

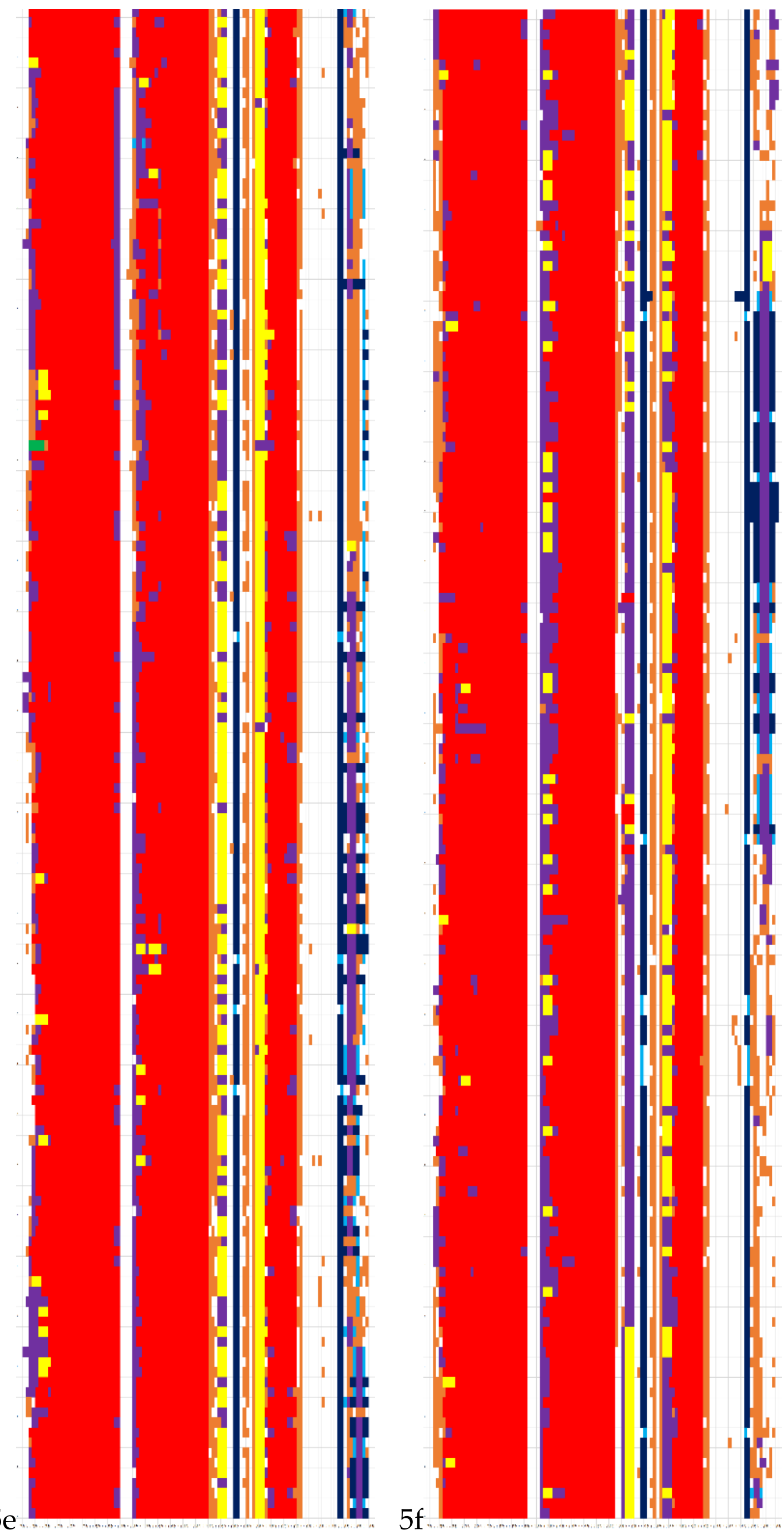

**Figure 5.** ***Secondary Structures of the WT at 350 K during 30 ns' MD simulations (up to down: seed1, seed2, seed3; left to right: neutral pH, low pH; 5a - neutral pH seed1, 5b - low pH seed1, 5c - neutral pH seed2, 5d - low pH seed2, 5e - neutral pH seed3, 5f - low pH seed3; for each of the six graphs its labels from up to down is 1 to 30 ns and its labels from right to left is PrP residue numbers 119-231).***

In Figure 6 we notice that at 450 K (i) under neutral pH environment, for seed1 the E β-sheet's two β-strands are broken, for seed2 the β-sheet structure becomes into β-bridge structure during 9-17.4 ns, and for all seeds the N- and C-terminals of $\alpha$1, $\alpha$2, $\alpha$3 are unfolding; and (ii) under low pH environment, β-sheet is broken for seed1, becomes into β-bridge for seed2, & is broken for seed3 from 18$^{th}$ ns, and for seed3 $\alpha$1 is unfolding from 20.6$^{th}$ ns, $\alpha$2 is unfolding from 26.6$^{th}$ ns (Figure 7) - we should also notice the RMSD values steadily go up under low pH environment at 450 K (Fig. 4.9a of [16]).

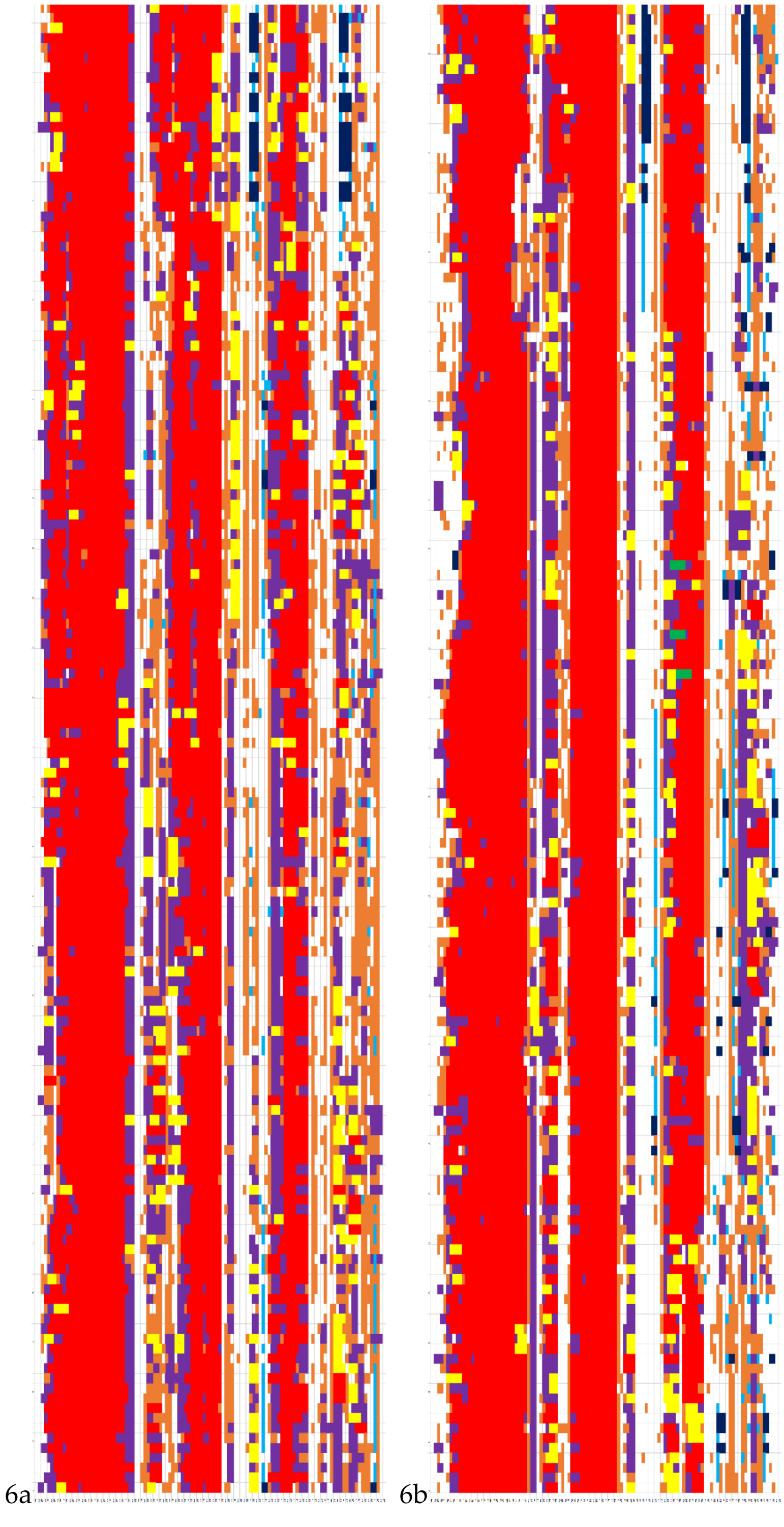

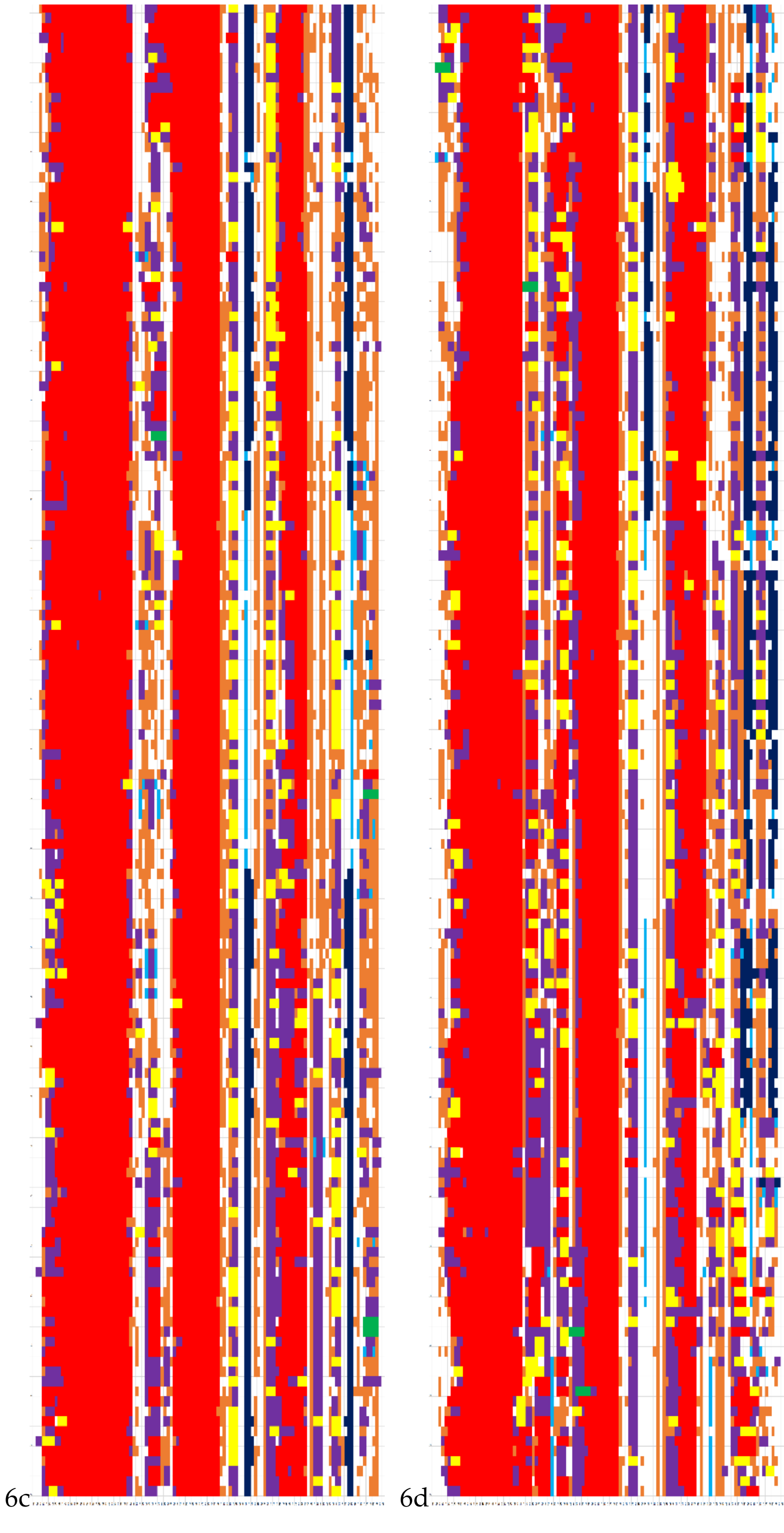

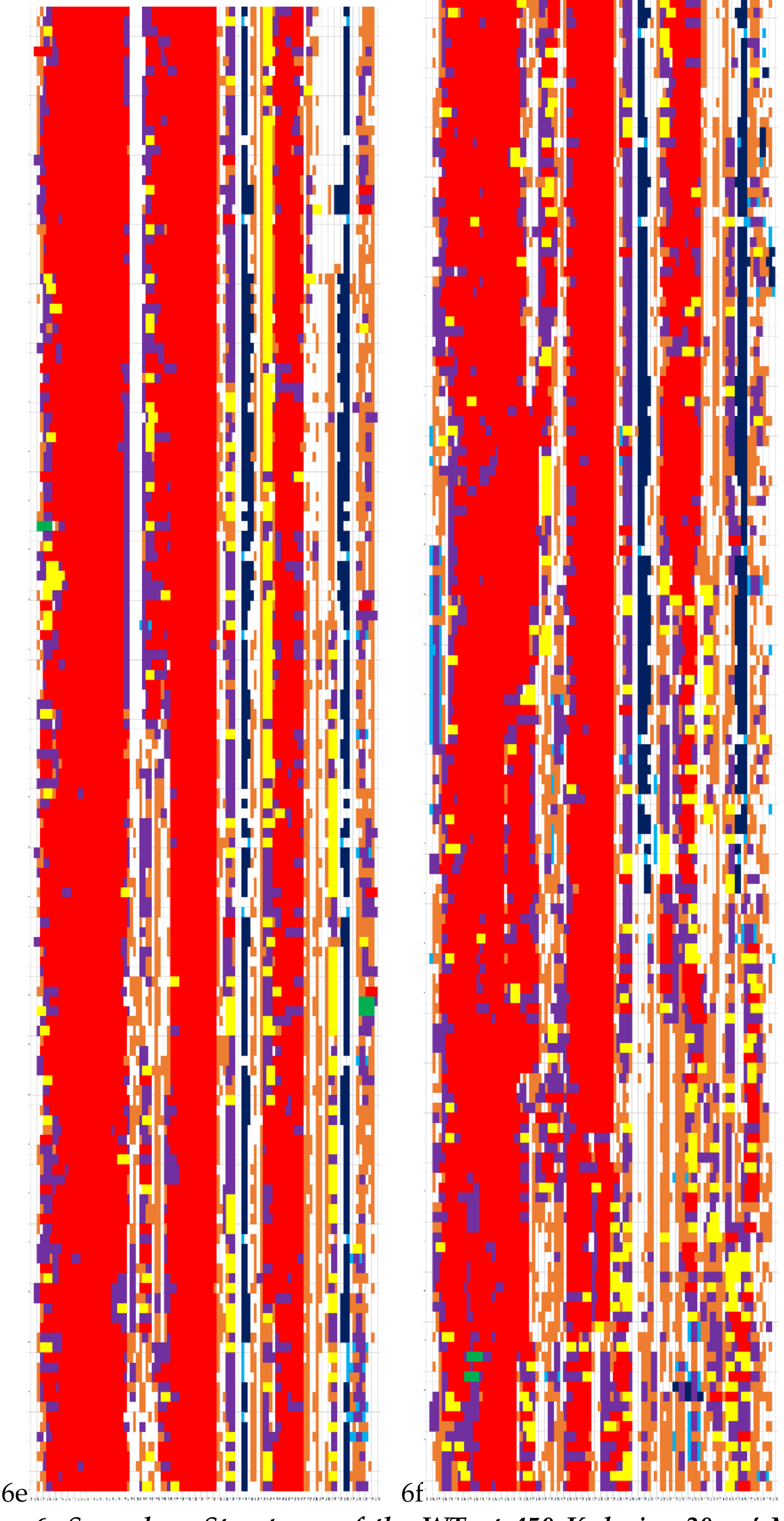

**Figure 6.** *Secondary Structures of the WT at 450 K during 30 ns' MD simulations (up to down: seed1, seed2, seed3; left to right: neutral pH, low pH; 6a - neutral pH seed1, 6b - low pH seed1, 6c - neutral pH seed2, 6d - low pH seed2, 6e - neutral pH seed3, 6f - low pH seed3; for each of the six graphs its labels from up to down is 1 to 30 ns and its labels from right to left is PrP residue numbers 119-231).*

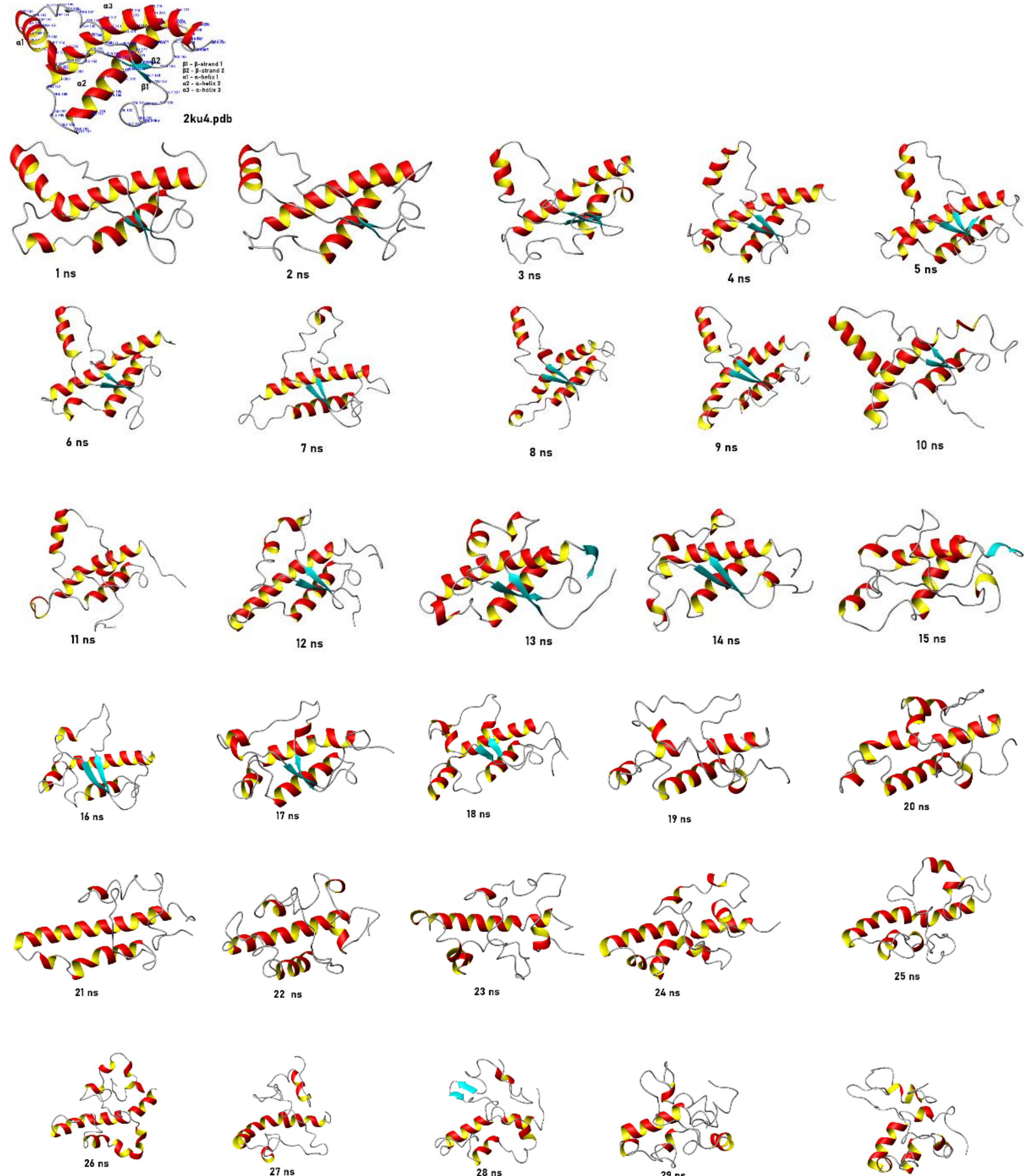

**Figure 7.** ***The 30 snapshots of three-dimensional molecular structures of horse PrP Wild-Type for the 6f – 30 ns' MD at 450 K low pH seed3: β-sheet is broken from 18$^{th}$ ns, and α1 is unfolding from 20.6$^{th}$ ns, α2 is unfolding from 26.6$^{th}$ ns.***

Generally, for WT, the above Secondary Structures results (shown in Table 2 and Figures 4-6), in combination with the RMSD (root mean square deviation), Radius of Gyration, RMSF (RMS fluctuation), B-factor results in Section 4.1 of [16], show us that at 450 K under low pH environment (compared with under neutral pH environment) $\alpha$-helices start to unfold and β-strands/bridges become more.

Let us do analyses on HBs, SBs, and HYDs of WT. Under low pH environment at MD-production constant temperatures of 300 K, 350 K and 450 K, all the HBs in Table 3 were removed. *Under neutral pH environment, at the temperature levels 300 K, 350 K and 450 K, the HBs ASP202-ARG+156/TYR157/TYR149 and ASP178-ARG+164/TYR128/TYR169 are always existing* (Table 3). Here we should denote that there exists a HB ASP178@O-TYR128@OH.HH 11.81 for seed1 at 350 K under low pH environment. In addition, in Table 3 the following HBs are also contributing to the structural stability of WT under neutral pH environment: GLU146-LYS+204 & GLN172-GLN219 at 300 K, GLU196-ARG+156 & GLU221-TYR163 & PRO158-ARG+136 at 350 K, and GLU196-GLY119/SER120 &

GLU221-SER167 & LEU138-TYR150 & GLY126/GLY127-ARG+164 & GLY131-GLN160 & GLU146-ARG+208 at 450 K.

**Table 3.** *All HBs (with % occupied rates >10% for seed1, seed2 and seed3) of WT at 300 K, 350 K and 450 K under neutral pH environment during the whole 3 sets of 30 ns' MD simulations for each temperature level:*

| *HBs at 300 K under neutral pH environment* | *HBs at 350 K under neutral pH environment* | *HBs at 450 K under neutral pH environment* |
|---|---|---|
| ASP202@OD1-TYR157@OH.HH 29.19, 84.64, 44.75 | GLU196@O-ARG$^+$156@NE.HE 10.27, 12.50, 0 | ASP202@OD1-ARG$^+$156@NE.HE 0, 0, 11.57 |
| ASP202@OD2-TYR157@OH.HH 62.79, 0, 49.53 | GLU196@O-ARG$^+$156@NH1.HH12 0, 0, 17.27 | ASP202@OD1-ARG$^+$156@NH1.HH12 23.27, 28.72, 14.80 |
| ASP202@OD1-ARG$^+$156@NH1.HH11 0, 67.38, 0 | GLU196@O-ARG$^+$156@NH2.HH21 21.24, 0, 0 | ASP202@OD1-ARG$^+$156@NH2.HH21 0, 5.00, 14.12 |
| ASP202@OD1-ARG$^+$156@NH1.HH12 60.02, 0, 42.64 | GLU196@O-ARG$^+$156@NH2.HH22 0, 0, 12.43 | ASP202@OD1-ARG$^+$156@NH2.HH22 16.40, 20.87, 16.53 |
| ASP202@OD2-ARG$^+$156@NH1.HH12 28.89, 0, 42.91 | GLU196@OE1-ARG$^+$156@NH2.HH21 0, 17.91, 0 | ASP202@OD2-ARG$^+$156@NE.HE 0, 0, 10.60 |
| ASP202@OD1-ARG$^+$156@NH2.HH22 33.65, 0, 48.65 | GLU196@OE2-ARG$^+$156@NE.HE 0, 10.62, 13.05 | ASP202@OD2-ARG$^+$156@NH1.HH12 17.80, 22.15, 17.12 |
| ASP202@OD2-ARG$^+$156@NH2.HH22 60.77, 0, 43.96 | GLU196@OE2-ARG$^+$156@NH2.HH21 0, 0, 11.42 | ASP202@OD2-ARG$^+$156@NH2.HH21 0, 0, 15.88 |
| ASP202@OD1-TYR149@OH.HH 0, 12.02, 0 | ASP202@OD1-TYR149@OH.HH 0, 19.46, 18.03 | ASP202@OD2-ARG$^+$156@NH2.HH22 21.58, 27.68, 13.57 |
| ASP202@OD2-TYR149@OH.HH 0, 67.42, 0 | ASP202@OD2-TYR149@OH.HH 0, 22.37, 41.93 | ASP202@OD1-TYR157@OH.HH 0, 20.78, 32.18 |
| ASP178@OD2-TYR128@OH.HH 27.04, 0, 0 | ASP202@OD1-ARG$^+$156@NE.HE 0, 0, 10.28 | ASP202@OD2-TYR157@OH.HH 0, 22.68, 29.00 |
| ASP178@OD1-TYR128@OH.HH 25.90, 0, 12.08 | ASP202@OD1-ARG$^+$156@NH1.HH11 0, 0, 45.58 | ASP202@OD1-TYR149@OH.HH 0, 7.25, 10.40 |
| ASP178@OD2-ARG$^+$164@NH2.HH21 0, 82.87, 40.21 | ASP202@OD1-ARG$^+$156@NH1.HH12 49.23, 24.93, 0 | GLU196@OE1-GLY119@N.H1 10.47, 0, 0 |
| ASP178@OD1-ARG$^+$164@NE.HE 0, 61.45, 30.03 | ASP202@OD1-ARG$^+$156@NH2.HH21 0, 0, 10.11 | GLU196@OE1-SER120@OG.HG 12.35, 0, 0 |
| ASP178@OD1-ARG$^+$164@NH2.HH21 0, 22.19, 30.73 | ASP202@OD1-ARG$^+$156@NH2.HH22 31.87, 28.07, 0 | ASP178@OD1-ARG$^+$164@NE.HE 0, 11.82, 8.88 |
| ASP178@OD2-ARG$^+$164@NE.HE 0, 0, 13.35 | ASP202@OD2-ARG$^+$156@NH1.HH11 0, 20.97, 0 | ASP178@OD1-ARG$^+$164@NH1.HH11 0, 0, 18.53 |
| GLU146@OE1-LYS$^+$204@NZ.HZ1 0, 16.53, 0 | ASP202@OD2-ARG$^+$156@NH1.HH12 28.89, 26.21, 0 | ASP178@OD1-ARG$^+$164@NH1.HH12 11.90, 0, 0 |
| GLU146@OE2-LYS$^+$204@NZ.HZ1 0, 16.10, 0 | ASP202@OD2-ARG$^+$156@NH2.HH21 0, 0, 14.52 | ASP178@OD1-ARG$^+$164@NH2.HH21 0, 17.60, 12.73 |
| GLU146@OE2-LYS$^+$204@NZ.HZ2 0, 16.09, 0 | ASP202@OD2-ARG$^+$156@NH2.HH22 45.29, 18.65, 0 | ASP178@OD2-ARG$^+$164@NH1.HH11 0, 0, 16.20 |
| GLU146@OE1-LYS$^+$204@NZ.HZ2 0, 15.53, 0 | ASP202@OD1-TYR157@OH.HH 43.71, 27.18, 66.71 | ASP178@OD2-ARG$^+$164@NH1.HH12 10.07, 0, 0 |
| GLU146@OE1-LYS$^+$204@NZ.HZ3 0, 17.57, 0 | ASP202@OD2-TYR157@OH.HH 30.47, 57.38, 0 | ASP178@OD2-ARG$^+$164@NH2.HH21 7.88, 18.63, 13.17 |
| GLU146@OE2-LYS$^+$204@NZ.HZ3 0, 17.39, 0 | ASP178@OD1-ARG$^+$164@NE.HE 13.48, 23.84, 15.06 | ASP178@OD1-TYR169@OH.HH 0, 27.60, 15.75 |
| GLN172@OE1-GLN219@NE2.HE21 10.89, 0, 0 | ASP178@OD1-ARG$^+$164@NH2.HH21 24.08, 15.08, 23.03 | ASP178@OD2-TYR169@OH.HH 0, 24.72, 17.75 |
| ASP178OD1-TYR169@HH 35.55, 0, 0 | ASP178@OD2-ARG$^+$164@NE.HE 36.29, 0, 0 | ASP178@OD1-TYR128@OH.HH 0, 12.25, 6.50 |
| ASP178OD2-TYR169@HH 40.47, 0, 0 | ASP178@OD2-ARG$^+$164@NH2.HH21 30.03, 23.94, 20.27 | ASP178@OD2-TYR128@OH.HH 0, 12.47, 5.90 |
| | ASP178@OD1-TYR169@OH.HH 32.27, 25.65, 57.75 | GLU221@OE1-SER167@OG.HG 11.70, 0, 0 |
| | ASP178@OD2-TYR169@OH.HH 0, 11.96, 30.44 | GLU221@OE2-SER167@OG.HG 10.38, 0, 0 |
| | GLU221@OE1-TYR163@OH.HH 23.40, 0, 26.15 | LEU138@O-TYR150@OH.HH 0, 10.75, 0 |
| | GLU221@OE2-TYR163@OH.HH 24.55, 14.65, 25.87 | GLY126@O-ARG$^+$164@NH2.HH21 0, 0, 15.48 |
| | PRO158@O-ARG$^+$136@NE.HE 0, 10.81, 0 | GLY127@O-ARG$^+$164@NE.HE 0, 0, 14.48 |
| | ASP178@OD1-TYR128@OH.HH 6.19, 35.51, 50.43 | GLY131@O-GLN160@NE2.HE21 0, 0, 13.70 |
| | ASP178@OD2-TYR128@OH.HH 5.35, 22.53, 38.97 | GLU146@OE1-ARG$^+$208@NE.HE 0, 0, 11.57 |
| | | GLU146@OE1-ARG$^+$208@NH2.HH21 5.88, 5.28, 11.72 |
| | | GLU146@OE2-ARG$^+$208@NH2.HH21 5.18, 0, 13.53 |

Table 4 lists all the SBs (with occupancy rates >5%) of the WT under neutral pH environment at 30 ns' MD-production constant temperatures of 300 K, 350 K and 450 K. All the SBs of 300 K (ASP144-ARG$^+$148, ASP147-HID$^+$140/ARG$^+$148/ARG$^+$151, HID177-LYS$^+$173, *ASP178-ARG$^+$164, HID187-ARG$^+$156*, GLU152-ARG$^+$148/ARG$^+$151, GLU168-ARG$^+$164, *GLU196-ARG$^+$156*/LYS$^+$194, *ASP202-ARG$^+$156*, GLU207-LYS$^+$204/ARG$^+$208, *GLU211-HID$^+$177*/ARG$^+$208/LYS$^+$220, GLU223-LYS$^+$220) are still existing at 350 K and 450 K. In addition, at 350 K there are additional SBs GLU223@CD-ARG$^+$228@CA.CZ & HID187@CG-ARG$^+$156@CA.CZ & ASP144@CG-HID$^+$140@ND1.HD1, and at 450 K there are additional SBs GLU146@CD-ARG$^+$208@CA.CZ, HID187@CG-LYS$^+$185@CA.NZ, GLU152@CD-ARG$^+$156@CA.CZ, HID187@NE2-LYS$^+$185@CA.NZ, GLU146@CD-HID$^+$140@ND1.HD1, ASP144@CG-HID$^+$140@ND1.HD1, HID140@NE2-ARG$^+$136@CA.CZ, ASP202@CG-LYS$^+$194@CA.NZ, GLU168@CD-ARG$^+$228@CA.CZ, GLU200@CD-HID$^+$187@ND1.HD1, GLU146@CD-LYS$^+$204@CA.NZ, HID140@NE2-ARG$^+$208@CA.CZ, HID140@CG-ARG$^+$136@CA.CZ, GLU200@CD-LYS$^+$194@CA.NZ, GLU221@CD-ARG$^+$228@CA.CZ, GLU207@CD-HID$^+$177@ND1.HD1. From Tables 3-4 we may know the polar contacts *ASP178-ARG$^+$164, ASP202-ARG$^+$156* (weaker), *GLU196-ARG$^+$156, GLU146-ARG$^+$208* (strong only at 450 K) contribute to the structural stability of WT during the MD simulations. The positions of the *special SBs* [HID187-ARG$^+$156 (linking $\alpha$2 and 310H1), GLU211-HID$^+$177 (linking $\alpha$3 and $\alpha$2)], the *special polar contacts* [ASP178-ARG$^+$164 (linking $\alpha$2 and β2-$3_{10}$H2-loop), ASP202-ARG$^+$156 (linking $\alpha$3 and $3_{10}$H1), GLU196-ARG$^+$156 (linking $\alpha$2-$\alpha$3-loop and $3_{10}$H1), GLU146-ARG$^+$208 (linking $\alpha$1 and $\alpha$3)], and *the SBs* [GLU223-LYS$^+$220, GLU207-LYS$^+$204/ARG$^+$208, GLU211-ARG$^+$208/LYS$^+$220, GLU223-LYS$^+$220 being within $\alpha$3, HID177-LYS$^+$173 being within $\alpha$2, ASP144-ARG$^+$148, ASP147-ARG$^+$148/ARG$^+$151 being within $\alpha$1, GLU196-LYS$^+$194 being within the $\alpha$2-$\alpha$3-loop, GLU152-ARG$^+$148/ARG$^+$151 linking the $\alpha$1-$3_{10}$H1-loop with $\alpha$1, GLU168-ARG$^+$164 linking the β2-$3_{10}$H2-loop and $3_{10}$H2, ASP147-HID$^+$140 linking $\alpha$1 and the β1-$\alpha$1-loop] should be noticed.

**Table 4.** *SBs (with occupancy rate >5%) of the WT during the three sets of 30 ns' MD simulations under neutral pH environment at 300 K, 350 K, 450 K (from left to right in turns: 300K seed1-seed3, 350 K seed1-seed3, 450 K seed1-seed3):*

| *SBs at 300 K under neutral pH environment* | *SBs at 350 K under neutral pH environment* | *SBs at 450 K under neutral pH environment* |
|---|---|---|
| ASP147@CG-ARG$^+$148@CA.CZ 100, 100, 100 | ASP147@CG-ARG$^+$148@CA.CZ 100, 100, 100 | ASP147@CG-ARG$^+$148@CA.CZ 100, 100, 100 |
| GLU211@CD-ARG$^+$208@CA.CZ 99.99, 98.53, 99.88 | GLU211@CD-ARG$^+$208@CA.CZ 99.75, 99.49, 99.19 | GLU211@CD-ARG$^+$208@CA.CZ 98.50, 99.32, 97.43 |
| GLU207@CD-LYS$^+$204@CA.NZ 99.70, 92.81, 99.90 | GLU207@CD-LYS$^+$204@CA.NZ 99.58, 95.18, 94.64 | GLU207@CD-LYS$^+$204@CA.NZ 81.43, 95.43, 95.37 |
| GLU221@CD-LYS$^+$220@CA.NZ 99.18, 95.05, 99.53 | GLU221@CD-LYS$^+$220@CA.NZ 95.19, 75.51, 95.72 | GLU152@CD-ARG$^+$148@CA.CZ 79.35, 37.07, 30.50 |
| GLU207@CD-ARG$^+$208@CA.CZ 97.93, 89.95, 90.66 | GLU223@CD-LYS$^+$220@CA.NZ 84.79, 88.04, 89.61 | GLU223@CD-LYS$^+$220@CA.NZ 52.90, 84.60, 85.40 |
| GLU223@CD-LYS$^+$220@CA.NZ 83.49, 86.61, 84.31 | HID177@CG-LYS$^+$173@CA.NZ 89.28, 68.55, 83.69 | GLU207@CD-ARG$^+$208@CA.CZ 78.65, 77.78, 71.73 |
| HID177@CG-LYS$^+$173@CA.NZ 76.03, 54.27, 59.23 | HID177@NE2-LYS$^+$173@CA.NZ 84.88, 59.56, 79.02 | GLU152@CD-ARG$^+$151@CA.CZ 64.87, 41.73, 48.42 |
| ASP178@CG-ARG$^+$164@CA.CZ 8.54, 80.97, 72.61 | GLU207@CD-ARG$^+$208@CA.CZ 82.31, 75.41, 70.49 | GLU221@CD-LYS$^+$220@CA.NZ 59.77, 67.45, 71.52 |
| ASP144@CG-ARG$^+$148@CA.CZ 56.94, 63.26, 81.40 | ASP144@CG-ARG$^+$148@CA.CZ 69.22, 66.07, 69.63 | HID177@CG-LYS$^+$173@CA.NZ 35.13, 61.08, 61.32 |
| HID187@NE2-ARG$^+$156@CA.CZ 75.33, 79.47, 78.88 | GLU196@CD-LYS$^+$194@CA.NZ 80.79, 46.75, 13.83 | HID177@NE2-LYS$^+$173@CA.NZ 24.10, 50.62, 48.95 |
| HID177@NE2-LYS$^+$173@CA.NZ 73.13, 44.92, 48.37 | GLU152@CD-ARG$^+$148@CA.CZ 19.88, 51.91, 50.99 | ASP147@CG-HID$^+$140@ND1.HD1 57.98, 56.63, 71.32 |
| ASP147@CG-HID$^+$140@ND1.HD1 32.85, 59.40, 5.03 | HID187@NE2-ARG$^+$156@CA.CZ 8.05, 50.02, 68.03 | ASP147@CG-ARG$^+$151@CA.CZ 53.13, 57.55, 52.87 |
| GLU152@CD-ARG$^+$148@CA.CZ 41.30, 57.81, 22.23 | ASP147@CG-HID$^+$140@ND1.HD1 66.73, 14.95, 50.97 | ASP144@CG-ARG$^+$148@CA.CZ 46.73, 40.42, 42.32 |
| GLU152@CD-ARG$^+$151@CA.CZ 40.62, 24.43, 46.85 | ASP147@CG-ARG$^+$151@CA.CZ 40.29, 51.06, 19.25 | GLU168@CD-ARG$^+$164@CA.CZ 18.28, 42.77, 9.27 |
| ASP147@CG-ARG$^+$151@CA.CZ 24.14, 6.27, 36.33 | GLU168@CD-ARG$^+$164@CA.CZ 38.75, 33.56, 57.47 | GLU146@CD-ARG$^+$208@CA.CZ 5.67, 19.18, 13.50 |
| ASP178@CG-HID$^+$177@ND1.HD1 5.60, 14.60, 21.83 | GLU152@CD-ARG$^+$151@CA.CZ 65.20, 42.93, 42.81 | ASP178@CG-HID$^+$177@ND1.HD1 16.07, 17.85, 19.48 |
| GLU211@CD-HID$^+$177@ND1.HD1 20.18, 8.37, 12.57 | GLU196@CD-ARG$^+$156@CA.CZ 17.17, 41.71, 43.47 | GLU211@CD-HID$^+$177@ND1.HD1 16.42, 16.27, 12.23 |
| GLU196@CD-LYS$^+$194@CA.NZ 0, 63.33, 0 | GLU211@CD-HID$^+$177@ND1.HD1 27.48, 15.45, 17.77 | HID187@CG-LYS$^+$185@CA.NZ 38.85, 5.92, 0 |
| GLU168@CD-ARG$^+$164@CA.CZ 52.00, 0, 8.60 | ASP178@CG-HID$^+$177@ND1.HD1 10.42, 14.12, 8.83 | ASP178@CG-ARG$^+$164@CA.CZ 0, 22.00, 38.58 |
| GLU196@CD-ARG$^+$156@CA.CZ 0, 9.63, 0 | ASP178@CG-ARG$^+$164@CA.CZ 10.45, 33.61, 0 | GLU152@CD-ARG$^+$156@CA.CZ 34.87, 0, 0 |
| ASP202@CG-ARG$^+$156@CA.CZ 0, 6.53, 0 | GLU223@CD-ARG$^+$228@CA.CZ 0, 0, 10.11 | HID187@NE2-LYS$^+$185@CA.NZ 25.88, 0, 0 |
| | HID187@CG-ARG$^+$156@CA.CZ 0, 0, 7.57 | GLU196@CD-LYS$^+$194@CA.NZ 0, 24.88, 14.97 |
| | ASP202@CG-ARG$^+$156@CA.CZ 0, 0, 6.69 | HID187@NE2-ARG$^+$156@CA.CZ 0, 15.08, 24.72 |
| | ASP144@CG-HID$^+$140@ND1.HD1 7.40, 0, 0 | GLU146@CD-HID$^+$140@ND1.HD1 24.42, 0, 16.37 |
| | | ASP144@CG-HID$^+$140@ND1.HD1 0, 5.23, 19.77 |
| | | GLU196@CD-ARG$^+$156@CA.CZ 0, 16.48, 16.42 |
| | | HID140@NE2-ARG$^+$136@CA.CZ 14.30, 0, 0 |
| | | ASP202@CG-LYS$^+$194@CA.NZ 0, 12.78, 14.15 |
| | | GLU168@CD-ARG$^+$228@CA.CZ 9.57, 0, 0 |
| | | ASP202@CG-ARG$^+$156@CA.CZ 9.35, 0, 0 |
| | | GLU200@CD-HID$^+$187@ND1.HD1 8.30, 0, 0 |
| | | GLU146@CD-LYS$^+$204@CA.NZ 0, 5.88, 5.47 |
| | | HID140@NE2-ARG$^+$208@CA.CZ 0, 0, 11.40 |
| | | HID140@CG-ARG$^+$136@CA.CZ 7.42, 0, 0 |
| | | GLU200@CD-LYS$^+$194@CA.NZ 0, 0, 10.70 |
| | | GLU221@CD-ARG$^+$228@CA.CZ 7.33, 0, 0 |
| | | GLU207@CD-HID$^+$177@ND1.HD1 6.42, 5.13, 0 |

Table 5 lists the basic HYDs (with occupancy rate 100%) maintained all the time under low or neutral pH environments at 300 K or 350 K or 450 K – all these basic HYDs are always contributing to the structural stability of the WT.

**Table 5.** *HYDs (with occupancy rate 100%) of the WT at 300 K, 350 K, 450 K during 30 ns' MD simulations whether under low or neutral pH environments:*

| *Under low pH environment* | *Under neutral pH environment* | *Position in the PrP structure* |
|---|---|---|
| PHE225@CB-ALA224@CA.C | PHE225@CB-ALA224@CA.C | Within α3 |
| ALA224@CB-PHE225@CA.C | ALA224@CB-PHE225@CA.C | Within α3 |
| VAL210@CB-VAL209@CA.C | VAL210@CB-VAL209@CA.C | Within α3 |
| VAL209@CB-VAL210@CA.C | VAL209@CB-VAL210@CA.C | Within α3 |
| VAL209@CB-MET206@CA.C | VAL209@CB-MET206@CA.C | Within α3 |
| MET206@CB-ILE205@CA.C | MET206@CB-ILE205@CA.C | Within α3 |
| ILE205@CB-MET206@CA.C | ILE205@CB-MET206@CA.C | Within α3 |
| VAL176@CB-PHE175@CA.C | VAL176@CB-PHE175@CA.C | Within α2 |
| PHE175@CB-VAL176@CA.C | PHE175@CB-VAL176@CA.C | Within α2 |
| VAL166@CB-PRO165@CA.C | VAL166@CB-PRO165@CA.C | Within β2-3$_{10}$H2-loop |
| PRO165@CB-VAL166@CA.C | PRO165@CB-VAL166@CA.C | Within β2-3$_{10}$H2-loop |
| ILE139@CB-LEU138@CA.C | ILE139@CB-LEU138@CA.C | Within β1-α1-loop |
| LEU138@CB-ILE139@CA.C | LEU138@CB-ILE139@CA.C | Within β1-α1-loop |
| LEU138@CB-PRO137@CA.C | LEU138@CB-PRO137@CA.C | Within β1-α1-loop |
| PRO137@CB-LEU138@CA.C | PRO137@CB-LEU138@CA.C | Within β1-α1-loop |
| MET134@CB-ALA133@CA.C | MET134@CB-ALA133@CA.C | Linking β1 and β1-α1-loop |
| ALA133@CB-MET134@CA.C | ALA133@CB-MET134@CA.C | Linking β1 and β1-α1-loop |
| LEU130@CB-MET129@CA.C | LEU130@CB-MET129@CA.C | Within β1 |
| MET129@CB-LEU130@CA.C | MET129@CB-LEU130@CA.C | Within β1 |
| VAL122@CB-VAL121@CA.C | VAL122@CB-VAL121@CA.C | Within N-terminal |
| VAL121@CB-VAL122@CA.C | VAL121@CB-VAL122@CA.C | Within N-terminal |
| MET213@CB-VAL210@CA.C 99.98, 100, 99.98 | MET213@CB-VAL210@CA.C 99.97, 100, 100 | Within α3 |
| MET206@CB-VAL203@CA.C 95.20, 100, 99.93 | MET206@CB-VAL203@CA.C 100, 99.90, 100 | Within α3 |

For HYDs, we also find some special and important HYDs listed in Table 6, which disappeared under low pH environment (except for the HYD VAL210@CB-VAL180@CA.C at 300 K under low pH environment), and completely disappeared at higher temperature 450 K. From Table 6, we know that the HYD VAL176@CB-ILE215@CA.C contributes the stability at 300 K under neutral pH environment, the HYD VAL210@CB-VAL180@CA.C contributes the stability at 300 K under low pH environment, and HYDs VAL210@CB-VAL180@CA.C, MET213@CB-VAL161@CA.C, VAL161@CB-MET213@CA.C not only contribute to the stability under neutral pH environment at 300 K but also at 350 K. All these *100%* HYDs disappeared at 450 K. Generally, we can see that low pH environment or/and higher temperature will make the protein structural stability become weaker.

**Table 6.** ***Some special HYDs (almost with occupancy rate 100%) of the WT at 300 K and 350 K during 30 ns' MD simulations under neutral pH environment:***

| *300 K – under neutral pH environment* | *350 K – under neutral pH environment* | | *Position in the PrP structure* |
|---|---|---|---|
| VAL210@CB-VAL180@CA.C | VAL210@CB-VAL180@CA.C | *100%* Exist at 300 K - low pH | Linking α3 and α2 |
| MET213@CB-VAL161@CA.C | MET213@CB-VAL161@CA.C | | Linking α3 and β2 |
| VAL161@CB-MET213@CA.C | VAL161@CB-MET213@CA.C | | Linking β2 and α3 |
| VAL176@CB-ILE215@CA.C | | | Linking α2 and α3 |

In summary, the molecular structure of the Mutant (compared with the WT) is unstable especially in the regions of $\alpha$2 & $\alpha$3 (especially at both terminals of $\alpha$2) and S167 is critical to the contribution of WT horse $PrP^C$ structure stability. The S167D mutation made the WT lost its *(i)* SBs such as ASP147-ARG148 (in $\alpha$1), ASP147-ARG151 (in $\alpha$1), ASP178-HIS177 (in $\alpha$2), ASP202-ARG156 (linking $\alpha$3 - $3_{10}$H1), GLU152-ARG151 (linking $\alpha$1-$3_{10}$H1-loop - $\alpha$1), GLU168-ARG164 (linking $3_{10}$H2 - β2-$3_{10}$H2-loop), GLU211-ARG208 (in $\alpha$3); *(ii)* an important polar contact ASP202-ARG$^+$156; *(iii)* two $\pi$-cations PHE141-ARG208.NH2$^+$ (linking β1-$\alpha$1-loop - $\alpha$3), HIS177-LYS173.NZ$^+$ (in $\alpha$2); *(iv)* one important HB GLU221-SER167; and redistributed the negative charges on the surface around the β2-$\alpha$2-loop region so that the well-defined and highly ordered β2-$\alpha$2-loop structure of WT horse $PrP^C$ has more variations in the Mutant. Our WT MD results have confirmed that the single amino acid differences at position 167 might influence the overall protein structures of WT.

## Conclusions

In [12], we were told the NMR structure of horse PrPC at 25 degrees C contains a well-structured and highly structure-ordered β2-$\alpha$2-loop, with $\alpha$3 this loop forms a binding site for a chaperone 'protein X', and within this loop the single amino acid S167 is unique to the PrP sequences of equine species. S167 was reported to be a protective residue for horse $PrP^C$. Our secondary structure WT MD studies show to us two performances *(i) for the WT the N-terminal half of $\alpha$1 is not stable and the β2-$\alpha$2-loop has less variations than other loops but (ii) from the optimized structures for the Mutant (compared with the WT) is unstable in the regions of $\alpha$2 & $\alpha$3 (especially at both terminals of $\alpha$2) and in the binding site for 'protein X'.* Detailed HBs, SBs, HYDs bioinformatics to explain the reasons for the performances were presented from the analyses of our WT MD results.

In summary, the author sets out to investigate the molecular differences between the wild-type (WT) horse prion protein and its S167D mutant which increases susceptibility to prion diseases, using molecular dynamics (MD). The author then claims that, based on the results the S167D mutation made some HBs & SBs in $\alpha$2 & $\alpha$3 and some $\pi$-cations lost, we may see the molecular structure of the Mutant (compared with the WT) is unstable especially in the regions of $\alpha$2 & $\alpha$3 (especially at both terminals of $\alpha$2) comparing to the WT. While purporting to elucidate the structural differences between the WT and mutant horse prion protein, this article reports no results from the MD simulations for the mutant, because currently the experimental structure of the mutant is not available yet. Various secondary structural statistics for the mutant simulations at different temperatures and pH conditions, and the changes in the H-bond network by making the mutation with MD simulations should be presented. Although the molecular mechanism of neurodegenerative diseases in the cases involving prion protein is still unclear, it is hypothesized to involve the growth of amyloid fibrils via the process of oligomerization (hence the higher beta-sheet population). Here we used a well folded helical protein monomer as the initial structure and runs MD simulations in a timescale (30 ns) much shorter than the one we required to observe significant protein conformational change; it is therefore still unclear how probative these simulations are of the biological process we try our best to model. However, the secondary structure time series plots (Figures 1-3) are highly effective providing us many representative structural clusters/snapshots to show us that for the WT under low pH condition it is much unstable than under neutral pH condition; though longer MD timescale is still needed to do furthermore. The simulations are very short (30 ns) to verify structural change. The simulation time should be increased to the conformational changes be noted as soon as the computing resources are available from NCI Australia.

For horse PrP S167D-mutant, currently there is not any literature data and not any experimental structure produced yet. But for horse $PrP^C$ WT, in Sections 5.2 and 5.3 of [19] and Section 4.1 of [16] there are rich bibliographic reviews and comparisons with rabbit $PrP^C$ and dog $PrP^C$, though updates are needed to be done.

This article is not to study the interactions of PrP with the solvent, the ions, and there is no ligand in this study for PrP binding with so that we did not include the interaction energies with the solvent (like [20] for example) and not calculate the energies using methods such as MM/PBSA. However, the free energy calculations are a research direction for this article to investigate the effects/contributions of ions such as $Cu^{2+}$ and of solvent such as waters. This should be highlighted as a furthermore future research direction for the author.